\documentclass[twocolumn,showpacs,amsmath,amssymb,prr,aps, superscriptaddress]{revtex4-1}
\usepackage[pdftex]{graphicx}
\usepackage{dcolumn}
\usepackage{bm}
\usepackage{units}
\usepackage{stmaryrd}
\usepackage[breaklinks=true,pdfborder={0 0 0}]{hyperref}
\usepackage{tabularx}

\usepackage{color}
\usepackage{tikz}
\usepackage{multirow}
\usepackage{braket}
\usepackage{siunitx}
\usepackage{makecell}
\usepackage{dsfont}
\usepackage{upgreek}

\newcommand{\circled}[1]{\raisebox{.5pt}{\textcircled{\raisebox{-.9pt} {\footnotesize #1}}}}

\makeatletter
\newcommand*{\encircled}[1]{\relax\ifmmode\mathpalette\@encircled@math{#1}\else\@encircled{#1}\fi}
\newcommand*{\@encircled@math}[2]{\@encircled{$\m@th#1#2$}}
\newcommand*{\@encircled}[1]{%
  \tikz[baseline,anchor=base]{\node[draw,circle,outer sep=0pt,inner sep=.2ex] {#1};}}
\makeatother

\renewcommand{\vec}[1]{\boldsymbol{\mathbf{#1}}}

\begin{document}
\title{Spin and orbital Edelstein effects in a two-dimensional electron gas: theory and application to AlO$_x$/SrTiO$_{3}$}

\author{Annika Johansson}
\email{annika.johansson@physik.uni-halle.de}
\affiliation{Institute of Physics, Martin Luther University Halle-Wittenberg,
06099 Halle, Germany}

\author{B{\"o}rge G{\"o}bel}
\affiliation{Institute of Physics, Martin Luther University Halle-Wittenberg,
06099 Halle, Germany}
\affiliation{Max Planck Institute of Microstructure Physics, Weinberg 2,
06120 Halle, Germany}

\author{J{\"u}rgen Henk}
\affiliation{Institute of Physics, Martin Luther University Halle-Wittenberg,
06099 Halle, Germany}

\author{Manuel Bibes}
\affiliation{Unit\'{e} Mixte de Physique, CNRS, Thales, Universit\'{e} Paris-Saclay, 91767, Palaiseau, France}

\author{Ingrid Mertig}
\affiliation{Institute of Physics, Martin Luther University Halle-Wittenberg,
06099 Halle, Germany}

\begin{abstract}
The Edelstein effect produces a homogeneous magnetization in nonmagnetic materials with broken inversion symmetry which is generated and tuned exclusively electrically. Often the spin Edelstein effect -- that is a spin density in response to an applied electric field -- is considered. In this Paper we report on the electrically induced magnetization that comprises contributions from the spin and the orbital moments. Our theory for these spin and orbital Edelstein effects is applied to the topologically nontrivial two-dimensional electron gas at the interface between the oxides SrTiO$_3$ and AlO$_x$. In this particular system the orbital Edelstein effect exceeds the spin Edelstein effect by more than one order of magnitude. This finding is explained  mainly by orbital moments of different magnitude in the Rashba-like-split band pairs, while the spin moments are of almost equal magnitude.
\end{abstract}



\maketitle

\section{Introduction}
One focus of the field of spintronics is to identify methods which allow to generate and manipulate spin-polarized electric currents efficiently, with the goal to realize powerful and nonvolatile electronic devices with reduced energy consumption~\cite{Wolf2001}. The initially proposed injection of spin-polarized currents from ferromagnets into semiconductors suffers from unavoidable inefficiency~\cite{Wolf2001, Schmidt2005}. To circumvent this shortcoming, spin-orbitronics aims at generating directly spin-polarized currents in pristine nonmagnetic materials~\cite{Manchon2015, Soumyanarayanan2016, Manchon2017}.

The perhaps most investigated phenomenon among the vast variety of spin-orbitronics effects is the spin Hall effect: a longitudinal charge current is accompanied by a transversal pure spin current or a spin voltage~\cite{Mott1965, Landau1965, Dyakonov1971, Hirsch1999, Kato2004_Science, Sinova2015}. Phenomenologically similar but of different physical origin is the Edelstein effect, also known as inverse spin-galvanic effect or current-induced spin polarization~\cite{Aronov1989, Edelstein1990, Gambardella2011}: in a pristine nonmagnetic system with broken inversion symmetry an applied electric field produces due to spin-orbit coupling a homogeneous spin density perpendicular to the field. A sizable number of systems have been identified which provide efficient charge-spin interconversion via the direct or the inverse Edelstein effect~\cite{Shen2014}: Rashba systems~\cite{Rashba1960, Rashba1984, Rashba1984_2}, semiconductors~\cite{Kato2004_PRL, Silov2004, Ganichev2006}, topological insulators~\cite{Culcer2010, Mellnik2010, Pesin2012, Ando2014, Li2014, Tian2015, Rojas2016, Zhang2016, Rodriguez2017}, Weyl semimetals~\cite{Johansson2018}, and oxide interfaces featuring two-dimensional electron gases (2DEGs)~\cite{Lesne2016, Seibold2017, Song2017, Vaz2019}. 

\begin{figure}
\includegraphics[width=0.5\textwidth]{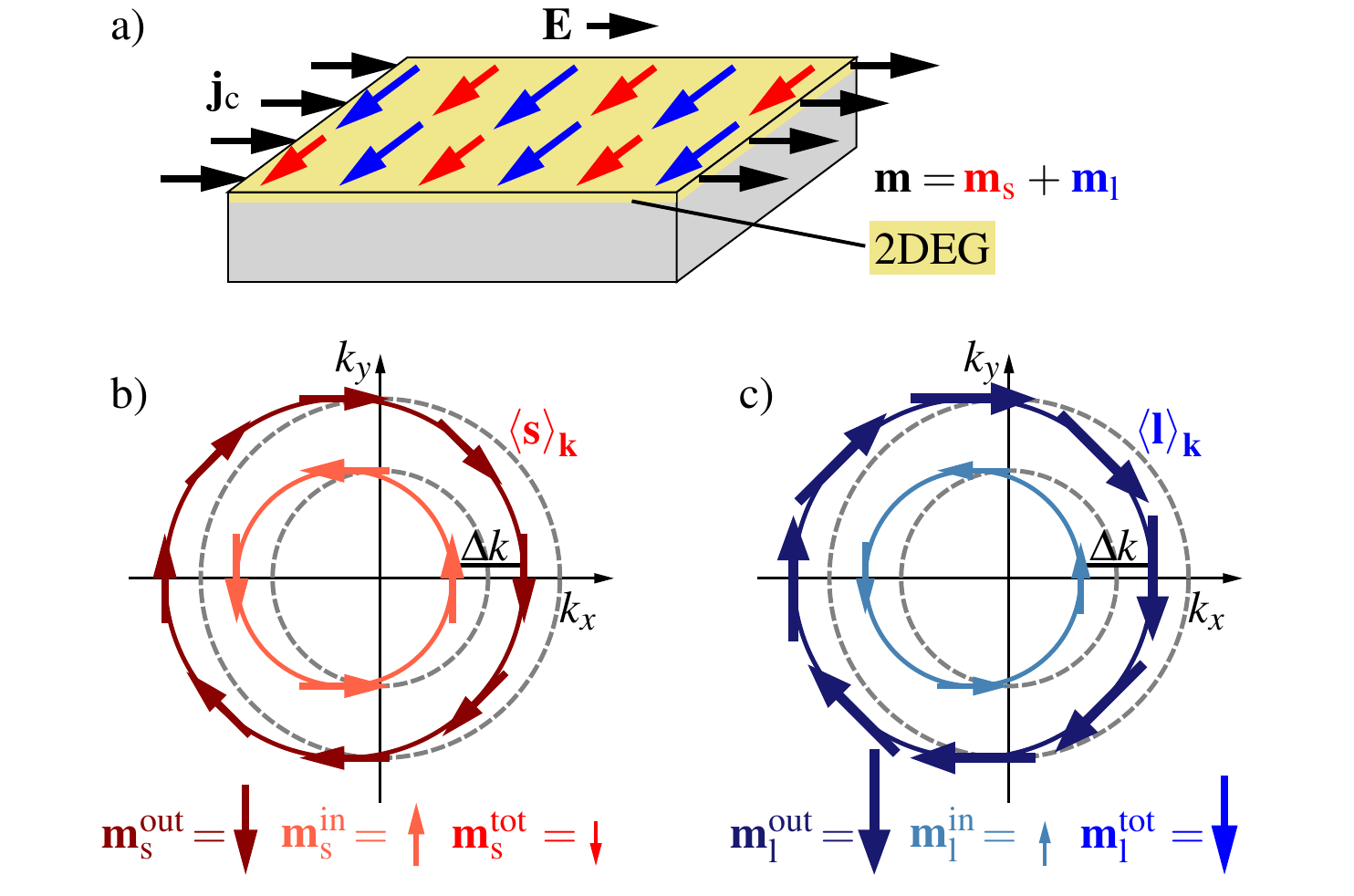}
\caption{Spin and orbital Edelstein effects illustrated for a Rashba-split 2DEG with spin- and orbital-momentum locking. An external electric field $\vec E$ induces a longitudinal charge current $\vec j_\text c$ as well as a homogeneous magnetization, usually perpendicular to $\vec E$. (a) Coexistence of spin ($\vec m_\mathrm{s}$, red) and orbital ($\vec m_\mathrm{l}$, blue) contributions to the total field-induced magnetic moment $\vec m$. (b) Spin Edelstein effect. Gray/red: equilibrium/nonequilibrium Fermi contours, arrows: spin expectation values. Due to the nonequilibrium redistribution of states, each Fermi contour provides a finite magnetization. Since these are of opposite sign the total magnetization is reduced. $\vec m_\text{s}$ is determined by the Rashba splitting $\Delta k$. (c) Orbital Edelstein effect. Similar to (b), but here the orbital moments (blue arrows) have different lengths, thereby reducing the compensation and eventually leading to a larger $\vec m_\mathrm{l}$.}
\label{img:EE_principle}
\end{figure}

In addition to their spin moment -- leading to the spin Hall and the spin Edelstein effect (SEE) -- electrons may also carry an orbital moment. Here, two contributions have to be distinguished: on the one hand the electrons' orbital motion and on the other hand the self-rotation of their wave packets~\cite{Chang1996}. These orbital moments can give rise to the orbital equivalents of the spin Hall and spin Edelstein effects, namely the orbital Hall~\cite{Zhang2005, Tanaka2008, Kontani2008, Kontani2009, Hanke2017, Jo2018} and the orbital Edelstein effect (OEE), the latter producing a current-induced orbital magnetization (Fig.~\ref{img:EE_principle}). 

Although the OEE has been predicted decades ago~\cite{Levitov1985}, it is often ignored when Edelstein effects are discussed. Only recently, the OEE resulting from the wave-packet self-rotation has been anticipated for helical and chiral crystals as well as for Rashba systems~\cite{Yoda2015, Yoda2018, Go2017}.  Furthermore, the SEE and OEE caused by the electrons' orbital motion have been discussed for noncentrosymmetric antiferromagnets~\cite{Salemi2019}. Both SEE and OEE in-plane responses perpendicular to the applied electric field $\vec{E}$ were found to be staggered, resulting in a zero net in-plane magnetization perpendicular to $\vec{E}$. However, nonzero field-induced magnetization components were predicted to exist parallel to the external field as well as out-of-plane. Importantly, the calculated orbital contribution to the Edelstein effect is larger than the spin contribution by at least one order of magnitude~\cite{Salemi2019}.

The above findings call for a theoretical investigation of the SEE and OEE in the 2DEG at the interface between AlO$_x$ (AO) and SrTiO$_3$ (STO), which does not show magnetic order in equilibrium. This particular system lends itself for such a study because of its promising properties concerning spintronics~\cite{Vaz2019}.

Using a semiclassical Boltzmann approach and an effective tight-binding model we calculate the SEE and OEE as responses to a static electric field. We predict a net OEE originating from the electrons' orbital motion that is more than one order of magnitude larger than its spin companion. Their dependence on the Fermi energy is traced back to band-resolved Edelstein signals. On top of this, we suggest  experiments to probe the large orbital contribution to the charge-magnetization interconversion, which is highly favorable for spin-orbitronic applications.
 
\section{2DEG at the AO/STO interface}
Although both SrTiO$_3$ and LaAlO$_3$ are three-dimensional (3D) bulk insulators, their two-dimensional (2D) interface features a conducting electron gas~\cite{Ohtomo2004}.  This 2DEG exhibits promising properties, such as high mobility~\cite{Ohtomo2004}, quantum transport~\cite{Ohtomo2004}, tunable carrier density and conductivity~\cite{Thiel2006, Cavaglia2008}, as well as highly efficient spin-to-charge conversion~\cite{Lesne2016, Seibold2017}. Recently, a 2DEG with similar properties has been found at the $\left(001\right)$ surface of STO covered by a thin Al layer~\cite{Roedel2016, Posadas2017}. This AO/STO 2DEG shows an inverse SEE of enormous magnitude, as is observed in a spin-pumping experiment~\cite{Vaz2019}. Its large signal is mainly caused by the interplay of spin-orbit coupling, the topological character of the 2DEG, and the high tunneling resistance of the AO layer.

An ideal STO bulk crystal has octahedral symmetry, its Fermi level lies within the fundamental band gap. Even in the presence of spin-orbit coupling the bands are twofold degenerate due to the coexistence of time-reversal and inversion symmetry~\cite{Zhong2013}. Interfaced with LAO or AO, however, the broken inversion symmetry lifts the spin degeneracy~\cite{Zhong2013, Khalsa2013}. Due to the interface constraint, bands originating from the $t_{2g}$ orbitals are shifted downwards in energy, thereby intersecting the Fermi level. The breaking of inversion symmetry allows for an additional mixing of orbitals (which is forbidden in the bulk) caused by spin-orbit coupling, thereby leading to a Rashba-like splitting of the bands~\cite{Zhong2013, Khalsa2013}.

For our investigation we use the effective eight-band tight-binding Hamiltonian proposed in Refs.~\onlinecite{Khalsa2013, Zhong2013, Vivek2017, Vaz2019} to model the $t_{2g}$ bands relevant for the formation of the 2DEG at the AO/STO interface. For details of the tight-binding model and its parameters~\cite{Vivek2017, Vaz2019} see Appendix~\ref{sec:hamiltonian}. 

\begin{figure}
\includegraphics[width=0.5\textwidth]{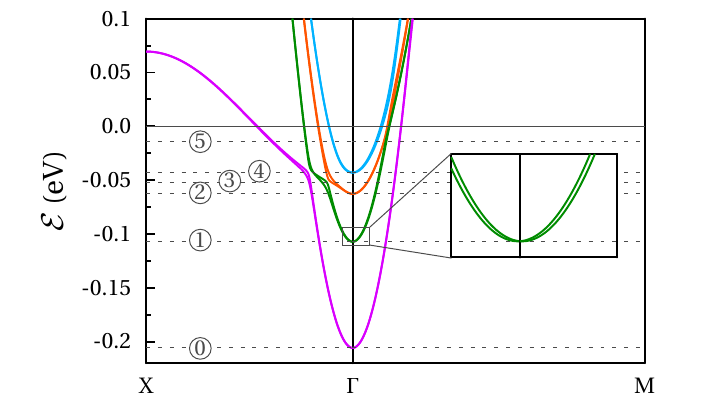}
\caption{Band structure of the AO/STO 2DEG computed within the tight-binding model. Selected energies are labeled as well as marked by dashed lines: $\circled{0}$ band edge of bands 1+2 ($\SI{-205}{\milli\electronvolt}$), \circled{$1$}  band edge of bands 3+4 ($\SI{-106}{\milli\electronvolt}$), $\circled{2}$  band edge of bands 5+6 ($\SI{-63}{\milli\electronvolt}$), $\circled{3}$  trivial avoided crossing ($\SI{-53}{\milli\electronvolt}$), $\circled{4}$  band edge of bands 7+8 ($\SI{-43}{\milli\electronvolt}$), and $\circled{5}$  topological band inversion ($\SI{-14}{\milli\electronvolt}$). The inset shows the Rashba-like splitting at $\circled{1}$.}
\label{img:band_structure}
\end{figure}

In the energy range around the Fermi level there exist four band pairs: two $d_{xy}$, one $d_{yz}$ and one $d_{zx}$ pair. These band pairs are identified in the band structure (Fig.~\ref{img:band_structure}). Spin-orbit coupling lifts the twofold spin degeneracy, leading to a Rashba-like splitting near the band edges (inset) and to avoided crossings between the second and third band pairs around $-53 \si{\milli\electronvolt}$ (label $\encircled{3}$). Around $-14 \si{\milli\electronvolt}$ ($\encircled{5}$) a topological band inversion (avoided crossing between the second and the third band pair) occurs, which is accompanied by one-dimensional spin-polarized helical edge states~\cite{Vivek2017, Vaz2019}.

\section{Results: Edelstein effect in the AO/STO 2DEG}
The current-induced magnetic moment $\vec m$ per 2D unit cell is calculated in the linear-response regime,
\begin{equation}\label{eq:induced_m}
\frac{A_0}{A} \vec m = \underbrace{\left( \chi^\mathrm{s} + \chi^\mathrm{l} \right)}_{\equiv \chi}  \vec E
\end{equation}
($A_0$ area of the unit cell, $A$ area of the entire system, $\vec E$ applied electric field). The rank-$2$ tensors $\chi^\mathrm{s}$ and $\chi^\mathrm{l}$ represent the conversion efficiencies of the SEE (s) and the OEE (l), respectively. Within the semiclassical Boltzmann approach the elements of these tensors read
\begin{equation}\label{eq:chi_boltzmann}
\begin{split}
\chi^\mathrm{s}_{ij} &= 2 \frac{A_0 e \mu_\text B}{A \hbar} \sum \limits_{\vec k} \braket{s}_{\vec k}^i \Lambda_{\vec k} ^ j \,\delta \left( \mathcal E_{\vec k} - \mathcal E_\text F \right), \\
\chi^\mathrm{l}_{ij} &=  \frac{ A_0 e \mu_\text B}{A \hbar} \sum \limits_{\vec k} \braket{l}_{\vec k}^i \Lambda_{\vec k} ^ j \,\delta \left( \mathcal E_{\vec k} - \mathcal E_\text F \right)
\end{split}
\end{equation}
($i, j = x, y, z$, elementary charge $e$, Bohr magneton $\mu_\text B$, reduced Planck constant $\hbar$, crystal momentum $\hbar\vec k$). The $\delta$-distributions restrict the band energies $\mathcal E_{\vec k}$ to the Fermi energy $\mathcal E_\text F$.
The $\vec k$-dependent spin expectation value $\braket{\vec{s}}_{\vec k}$  and the orbital-momentum expectation value $\braket{\vec l}_{\vec k}$ are weighted with the mean free path $\vec{\Lambda}_{\vec k}$. The latter is given in constant relaxation-time approximation, $\vec \Lambda_{\vec k}= \tau_0 \vec v_{\vec k}$ with the relaxation time $\tau_0$ and the group velocity $\vec v_{\vec k} = \frac{1}{\hbar}\frac{\partial}{\partial \vec k} \mathcal E_{\vec k}$. The redistribution of the electronic states caused by the external electric field leads to a nonequilibrium magnetization that is attributed either to the spin or to the orbital moments (Fig.~\ref{img:EE_principle}b and c). Details of the Boltzmann approach are presented in Appendix~\ref{sec:boltzmann}.

The coexistence of time-reversal symmetry and mirror planes perpendicular to the $\left\langle 1 0 0 \right\rangle$ and $\left\langle  1 1 0 \right\rangle$ directions dictates that the spin as well as the orbital moments are oriented completely within the plane. Further, these symmetries allow only nonzero tensor elements $\chi_{xy}^{\cdot} = - \chi_{yx}^{\cdot}$. In particular the diagonal elements $\chi_{xx}^{\cdot}$ and $\chi_{yy}^{\cdot}$ as well as the out-of-plane elements $\chi_{zx}^{\cdot}$, $\chi_{zy}^{\cdot}$, $\chi_{xz}^{\cdot}$ and $\chi_{yz}^{\cdot}$, which are predicted in Ref.~\onlinecite{Salemi2019} for noncentrosymmetric antiferromagnets, are forbidden by the mirror symmetries in the AO/STO system.

In isotropic Rashba systems [for example realized in the $L$-gap surface state in Au(111)~\cite{Henk2003, Hoesch2004, Henk2004}], the spin moments of the Rashba-split states are oriented antiparallel to each other with equal absolute values.  Thus, although each individual state would induce a pronounced Edelstein effect, the resulting total SEE from such a pair is strongly reduced due to partial compensation of the oppositely orientated moments, as illustrated in Fig.~\ref{img:EE_principle}(b).

In order to discuss the Edelstein effect in the AO/STO system, we define the $\vec{k}$-dependent quantities $\Delta k$, $\Delta \Lambda_{\vec{k}}$, $\Sigma_{\vec{s}}$, and $\Sigma_{\vec{l}}$. These quantities contain properties of two states $\vec{k}_1$ and $\vec{k}_2$ which would be degenerate in the absence of spin-orbit coupling. As discussed above, in a Rashba system these states' spin expectation values are oppositely oriented and of equal absolute value, which leads to a pronounced compensation of the Edelstein contributions. An increased Edelstein effect should show up if this compensation is diminished by
\begin{enumerate}
    \item a large $\Delta k=|k_1-k_2|$, which is due to large spin-orbit interaction and particularly shows up at avoided crossings,  accompanied by a large difference of the band-resolved densities of states,
    \item a large difference of mean free paths $\Delta \Lambda_{\vec{k}}=|\vec{\Lambda}_{\vec{k}_1}-\vec{\Lambda}_{\vec{k}_2}|$,
    \item a large sum $\Sigma_{\vec{s}}=\Braket{\vec{s}}_{\vec{k}_1}+\Braket{\vec{s}}_{\vec{k}_2}$ resp. $\Sigma_{\vec{l}}=\Braket{\vec{l}}_{\vec{k}_1}+\Braket{\vec{l}}_{\vec{k}_2}$, meaning that the spins or orbital moments of the Rashba-like-split states are not perfectly antiparallelly aligned or differ in absolute value.
\end{enumerate}
The first and second factor affect the SEE and OEE in equal measure, whereas the third may lead to a significantly different charge--magnetic-moment conversion, as sketched in Figs.~\ref{img:EE_principle}(b) and (c). The second point is less relevant for the present study, since we assume a constant relaxation time $\tau_0$.

In Rashba systems -- and consequently in the present tight-binding model -- spin-orbit coupling is necessary to lift spin degeneracy. Splitting the bands of a Kramers pair by $\Delta  k$ then produces nonzero spin and orbital moments $\Braket{\vec{s}}_{\vec{k}}$ and $\Braket{\vec{l}}_{\vec{k}}$. It turns out that hybridization among the set of $t_{2g}$ orbitals is crucial for the emergence of nonzero orbital moments. A pure $d_{xy}$/$d_{yz}$/$d_{zx}$ state would have vanishing expectation value, as is clear by representing the orbital moment $\hat{\vec{l}}$ in the $t_{2g}$ basis (Appendix~\ref{sec:hamiltonian}).

\begin{figure}
\includegraphics[width=0.5\textwidth]{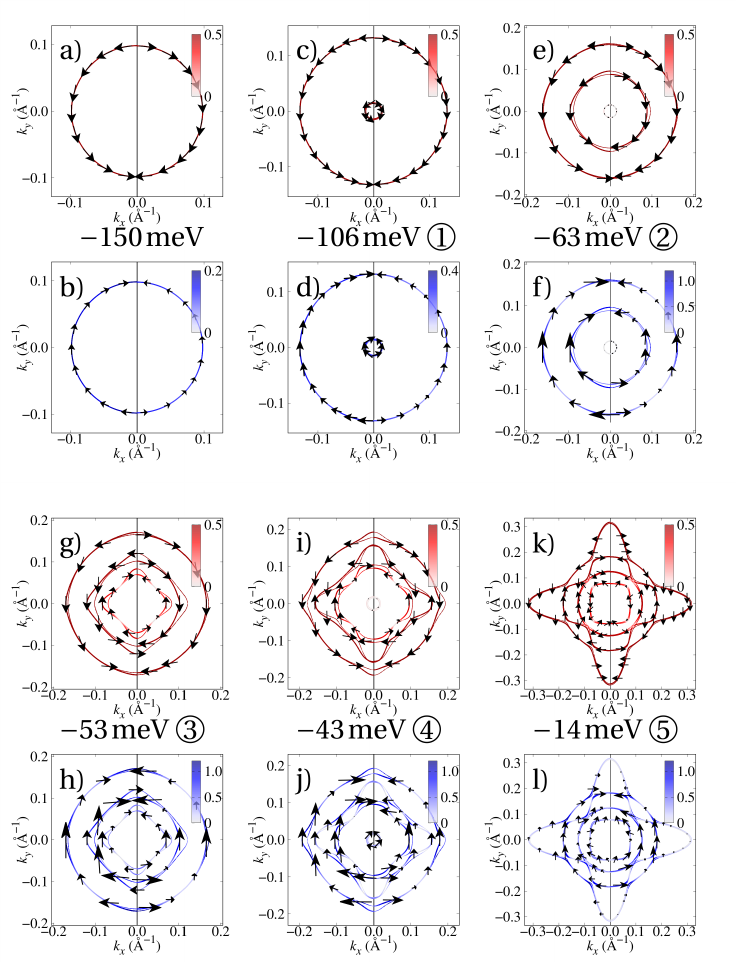}
\caption{Expectation values of spin (red) and orbital (blue) moments (in units of $\hbar$) in a AO/STO 2DEG at selected isoenergy contours, computed within the tight-binding model. Colors represent the absolute values of the moments. The arrows on the left (right) half of each panel depict the direction of the outer (inner) bands' moments. The encircled labels correspond to the energies indicated in Fig.~\ref{img:band_structure}, whereas panels c--f as well as i and j show the isoenergy contours slightly above the band edges.}
\label{img:spin_orbit}
\end{figure}

In Rashba systems with two parabolic bands (free electrons) the spin moments of both bands have the same absolute values, thus,  $\Sigma_{\vec{s}}$ vanishes. Consequently, the contributions of both bands to the SEE partially compensate. The outer band (with larger $k$) dominates and determines the sign of the SEE because of the higher density of states.  In the AO/STO 2DEG, however, the $\vec{k}$-resolved spin and orbital moments exhibit more complex textures than in the free-electron Rashba model, as is evident from  Fig.~\ref{img:spin_orbit}. Near the band edges the spin moments exhibit a texture close to that in the free-electron Rashba model -- that is equal absolute values for both bands of a pair [Figs.~\ref{img:spin_orbit}(a),  (c), and (e)]; on the contrary, the orbital moments of a band pair are also directed tangentially to their Fermi contour, but the absolute values differ [(b), (d), and (f)]. This variation is explained by the spin and magnetic quantum  numbers. A spin of $s = \nicefrac{1}{2}$ has only two quantum numbers ($m_s = \pm \nicefrac{1}{2}$), whereas for an orbital moment of $l = 2$ (d orbitals) five ($m_l = 0, \pm 1, \pm 2$) are allowed. This larger variety of quantum numbers is reflected in the expectation values  of the orbital moments $\Braket{\vec{l}}_{\vec{k}}$, their absolute values can considerably differ among a band pair.

Following up on the above, even a small $\Sigma_{\vec{l}}$ may produce a sizable OEE that could exceed the SEE [Figs.~\ref{img:EE_principle}(b) and (c)]: the unequal absolute values of $\Braket{\vec{l}}_{\vec{k}}$ reduce the compensation of oppositely oriented orbital moments and thereby lead to a more pronounced OEE.  As an example consider energies close to the band edge of the lowest band pair. There the OEE surpasses the SEE by a factor of $3$, although the absolute values of the orbital moments are merely $40 \%$ of the spin moments [Fig.~\ref{img:spin_orbit}(a) and (b)] and the mean difference of $|\Braket{\vec{l}}_{\vec{k}}|$ of this Kramers pair amounts to only $3\%$ of the average spin moment.

This conjecture holds also for energies for which the isoenergy contours deviate strongly from the circular ones of the free-electron Rashba model. As an example consider the contours chosen in  Figs.~\ref{img:spin_orbit}(g)--(l), whose spin and orbital moments exhibit complex textures. The nonzero $\Sigma_{\vec{s}}$ [(g), (i), and (k)] yields a sizable SEE. However, $\Sigma_{\vec{l}}$ is larger, that is why the OEE is larger as well.

\begin{figure}
\includegraphics[width=0.5\textwidth]{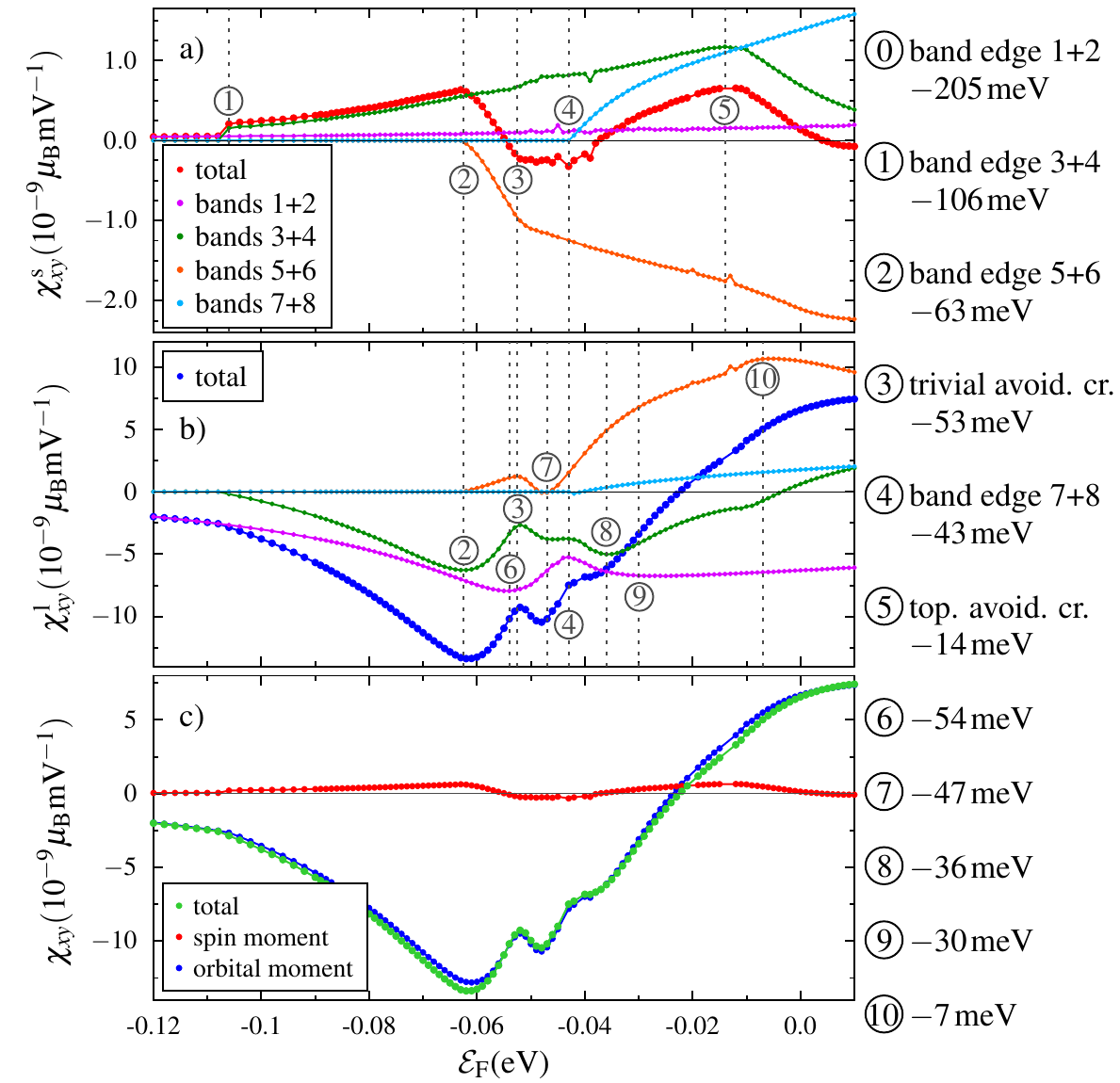}
\caption{Efficiency of the spin and orbital Edelstein effects in the AO/STO 2DEG calculated within the tight-binding model. The band-resolved spin (a) and orbital (b) contributions as well as the total Edelstein efficiency (c) are shown. Encircled labels indicate the selected energies listed on the right-hand side. The relaxation time is set to $\tau_0 = \SI{1}{\pico\second}$.}
\label{img:Edelstein_STO}
\end{figure}

The above examples suggest pronounced signatures of the energy dependencies of the Edelstein efficiencies $\chi_{xy}^\mathrm{s}$ and $\chi_{xy}^\mathrm{l}$ themselves (Fig.~\ref{img:Edelstein_STO}). The latter are understood in detail by resolving contributions of individual bands. 

It suffices to recapitulate briefly the energy dependence of $\chi_{xy}^\mathrm{s}$, depicted in  Fig.~\ref{img:Edelstein_STO}(a), since it has been discussed in detail in Ref.~\onlinecite{Vaz2019}. We focus on three effects: density of states, spin texture, and hybridization. 

A step-like increase in the total signal at $\encircled{1}$ in Fig.~\ref{img:Edelstein_STO}(a) indicates that the band edge of the second pair increases the number of contributing states. 
The third Kramers pair has reversed spin chirality with respect to the other pairs [Fig.~\ref{img:spin_orbit}(g)]. This feature explains the opposite sign of $\chi_{xy}^\mathrm{s}$ of this pair [orange in Fig.~\ref{img:Edelstein_STO}(a)] and the extremum of the total signal, clearly showing up at $\encircled{2}$. The reversal is compensated by the usual spin chirality of the fourth pair; confer for example $\encircled{4}$. Uncompensated spin textures produce extrema as well: the maximum at  $\encircled{5}$ is traced back to the topological band inversion [enhanced $\Sigma_{\vec{s}}$; shown in Fig.~\ref{img:spin_orbit}(k)].

Eventually, the signal scales with the splitting $\Delta k$ of a Kramers pair near band edges. $\Delta k$ is enhanced by orbital hybridization which  produces the extremum near the avoided crossing at $\encircled{3}$.

We now turn to the orbital efficiency $\chi_{xy}^\mathrm{l}$ whose magnitude is affected by the non-monotonic behavior of $\Delta k$, $|\Braket{\vec{l}_{\vec{k}}}|$, and particularly $\Sigma_{\vec{l}}$.  

As pointed out above, the band-resolved $|\Braket{\vec{l}}_{\vec{k}}|$'s of a Kramers pair differ in the free-electron-like regime, in contrast to the $|\Braket{\vec{s}}_{\vec k}|$'s.  This reduced compensation ($\Sigma_{\vec l} \neq 0$) leads in combination with $\Delta k$ to a significant but smooth increase of the current-induced magnetization. This signature is exemplified by 
$\chi_{xy}^\mathrm{l}$ shown in Fig.~\ref{img:Edelstein_STO}(b) for bands 1+2 ($\encircled{0} \ldots \encircled{6}$), bands 3+4 ($\encircled{1}$ and $\encircled{2}$), bands 5+6 ($\encircled{2}$ and $\encircled{3}$), and bands 7+8 (energies above $\encircled{4}$).

The clear-cut relation of spin chirality and sign of $\chi_{xy}^\text{s}$ established above does not hold for $\chi_{xy}^\text{l}$. Although all band pairs exhibit the same orbital chirality (Fig.~\ref{img:spin_orbit}), the signs of the band-pair-resolved $\chi_{xy}^\text{l}$ vary because of different distributions of $\Braket{\vec{l}}_{\vec{k}}$.
For example, the outer states of the pairs 1+2 and 3+4 have a larger $|\Braket{\vec{l}}_{\vec{k}}|$  [(b) and (d)] as well as a higher density of states than the inner ones; thus, these states determine the sign of the Edelstein signal. Likewise for the other band pairs: here, the `inner' $|\Braket{\vec{l}}_{\vec{k}}|$ are larger than `outer' ones, but despite their lower density of states they set the sign of  $\chi_{xy}^\mathrm{l}$ [(h) and (j)].

At elevated energies, the band-resolved efficiencies $\chi_{xy}^\mathrm{l}$ show a number of extrema. Those at $\encircled{3}$, $\encircled{4}$, and $\encircled{7}$ are related to $\Sigma_{\vec{l}}$, whereas those at $\encircled{8}$, $\encircled{9}$, and $\encircled{10}$ are associated with $\Delta k$.

The features of the individual bands are carried forward to the total OEE signal; this sum of all orbital contributions exhibits several extrema and a sign change around $\SI{-20}{\milli\electronvolt}$. Abrupt steps or sharp extrema do not occur in $\chi_{xy}^\mathrm{l}$ because the band-resolved contributions are smooth; this finding is at variance with that for $\chi_{xy}^\mathrm{s}$.

Avoided crossings -- either topologically trivial ($\encircled{3}$) or nontrivial ($\encircled{5}$) -- do not lead to pronounced extrema of the  orbital signal, although they do so in the spin signal. One consequence of the trivial avoided crossing is to increase $\Delta k$ of the second and third band pair. While this increased splitting implies an enhanced SEE, it does not affect the orbital signal significantly since the latter is partially compensated by the reduction of $\Sigma_{\vec{l}}$ [Fig.~\ref{img:spin_orbit}(h)].  Furthermore, $\chi_{xy}^\mathrm{l}$ does not exhibit an extremum at the topological avoided crossing, because there the emergence of uncompensated orbital moments ($\Sigma_{\vec{l}}$) is less pronounced than that of  uncompensated spin moments ($\Sigma_{\vec{s}}$). 

The total Edelstein efficiency $\chi_{xy}$ defined in Eq.~\eqref{eq:induced_m} is dominated by the orbital contribution which exceeds the spin contribution by one order of magnitude [Fig.~\ref{img:Edelstein_STO}(c)]. 
In other words, the fact that $\Sigma_{\vec{l}} > \Sigma_{\vec{s}}$ for bands of Kramers pair causes $|\chi^\text{l}_{xy}| > |\chi^\text{s}_{xy}|$. This implicates that for any experiment in which the total current-induced magnetization is measured both spin and orbital moments should be considered as sources of the observed Edelstein effect. This calls for experiments that are able to discriminate both origins. 

\section{Detection of the orbital Edelstein effect}\label{sec:experiment}
A vast number of experiments addresses the charge-spin interconversion via the direct or the inverse Edelstein effect; to name a few: Refs.~\onlinecite{Kato2004_PRL, Kato2004_Nature, Silov2004, Ganichev2006, Sih2005, Johnson1998, Hammar2000, Ando2014, Li2014, Tian2015, Rojas2013, Tserkovnyak2002, Rojas2016, Lesne2016, Song2017, Liu2011, Kondou2016, Vaz2019}. These are usually interpreted within a theoretical framework in which merely the spin contribution to the Edelstein effect is considered. However, as we have demonstrated in the previous section, the orbital contribution cannot be neglected \textit{a priori}.

In a system with broken inversion symmetry the magnetization induced by the direct Edelstein effect would be of both spin and orbital origin, unless s electrons ($l = 0$) dominate the transport. An optical experiment sensitive to the surface magnetization would detect both contributions, e.\,g.\ a magneto-optical Kerr experiment in longitudinal geometry. Furthermore, X-ray magnetic circular dichroism (XMCD) allows to distinguish the spin and the orbital magnetic moments via the so-called sum rules~\cite{Thole1992, vanderLaan1998}.

In spin-pumping experiments on the inverse Edelstein effect (e.\,g.\ Refs.~\onlinecite{Rojas2013, Rojas2016, Lesne2016, Vaz2019}) a pure spin current is injected into a two-dimensional system and converted there into a charge current. This pure spin current gives rise to an inverse SEE, but not to an inverse OEE. 

\section{Synopsis}
Our theoretical investigation of the spin and orbital Edelstein effects in the 2DEG at an AO/STO interface reveals that the contribution from the orbital moments exceeds that of the spin moments by more than one order of magnitude. This finding is explained  mainly by larger variations of the band-resolved orbital moments along the Fermi contour as compared to those of the spin moments. From this follows that the orbital contribution proves significant in applications which rely on the direct Edelstein effects. The orbital Edelstein effect could be distinguished from its spin counterpart by XMCD experiments.

Applications featuring an inverse Edelstein effect, e.\,g.\ the proposed magnetoelectric spin-orbit device~\cite{Manipatruni2019}, would be more efficient  by utilizing the inverse OEE, that is by injecting an orbital current in addition to a spin current. 

\acknowledgements
This work is supported by CRC/TRR $227$ of Deutsche Forschungsgemeinschaft (DFG) and the ERC Advanced grant number $833973$ ``FRESCO''. MB thanks the Alexander von Humboldt Foundation for supporting his stays at Martin Luther University Halle-Wittenberg.

\appendix

\section{Electronic-structure calculations}\label{sec:hamiltonian}
The relevant electronic states that form the 2DEG at the AO/STO interface are derived from four $t_{2g}$ orbitals~\cite{Khalsa2013, Zhong2013}. It is thus natural to utilize the  tight-binding Hamiltonian proposed in Refs.~\onlinecite{Khalsa2013, Zhong2013, Vivek2017} and its extension to eight bands proposed in Ref.~\onlinecite{Vaz2019}. 

The basis set consists of four spin-up orbitals -- $\{ d_ {xy \uparrow}^{(1)}, d_ {xy \uparrow}^{(2)}, d_ {yz \uparrow}, d_ {zx \uparrow} \}$ -- and four spin-down orbitals --  $\{ d_ {xy \downarrow}^{(1)}, d_ {xy \downarrow}^{(2)}, d_ {yz \downarrow}, d_ {zx \downarrow} \}$. The superscripts $1$ and $2$ for the $d_{xy}$ orbitals account for  crystal-field splitting; see Eq.~\eqref{eq:model_parameters} below. The Hamiltonian matrix
\begin{equation}\label{eq:TB_Hamiltonian}
\hat{H} = 
\begin{pmatrix}
H^+ & H_\lambda \\
H_\lambda^\dagger & H^-
\end{pmatrix}
\end{equation}
decomposes thus into the blocks
\begin{equation}\label{eq:H_g}
H^\pm =
\begin{pmatrix}
\mathcal E_{xy}^{(1)} & 0 & \mathrm i g_1 \sin \tilde k_x & \mathrm i g_1 \sin \tilde{k}_y \\
0 & \mathcal E_{xy}^{(2)} & \mathrm i g_2 \sin \tilde k_x & \mathrm i g_2 \sin \tilde{k}_y \\
- \mathrm i g_1 \sin \tilde k_x & - \mathrm i g_2 \sin \tilde k_x & \mathcal E_{yz} & \pm \mathrm i \lambda \\
-\mathrm i g_1 \sin \tilde k_y & - \mathrm i g_2 \sin \tilde k_y & \mp \mathrm i \lambda & \mathcal E_{zx}
\end{pmatrix}
\end{equation}
and
\begin{equation}\label{eq:H_lambda}
H_\lambda = \lambda   
\begin{pmatrix}
0 & 0 & 1 & - \mathrm i \\
0 & 0 & 1 & - \mathrm i \\
-1 & -1 & 0 & 0 \\
\mathrm i & \mathrm i & 0 & 0 
\end{pmatrix}
\end{equation}
with $\tilde k_i = k_i \, a$ ($a$ lattice constant). The diagonal terms \begin{equation}\label{eq:diagonal}
\begin{split}
\mathcal E_{xy}^{(i)}&=2t \left(2 - \cos \tilde k_x - \cos \tilde k_y \right) + \mathcal E_{0xy}^{(i)}, \quad i = 1, 2, \\
\mathcal E_{yz}&=2t \left(1-\cos \tilde k_y \right) + 2 t_h \left(1- \cos \tilde k_x \right) + \mathcal E_{0z}, \\
\mathcal E_{zx}&=2t \left(1-\cos \tilde k_x \right) + 2 t_h \left(1- \cos \tilde k_y \right) + \mathcal E_{0z},
\end{split}
\end{equation}
reflect the band structure without spin-orbit coupling. $t$ and $t_h$ denote the strength of nearest-neighbor hopping of the light and the heavy bands, respectively. Terms proportional to $\lambda$ mimic atomic spin-orbit coupling, while those proportional to $g_1$ or $g_2$ account for the interatomic and orbital-mixing spin-orbit interaction, which stems from the deformation of the orbitals at the interface~\cite{Petersen2000, Khalsa2013, Zhong2013}.
The parameters for the AO/STO interface, \begin{equation}\label{eq:model_parameters}
\begin{array}{l l l l}
\mathcal E_{0xy}^{(1)}&=\SI{-205}{\milli\electronvolt}, & t&=\SI{388}{\milli\electronvolt}, \\
\mathcal{E}_{0xy}^{(2)}&=\SI{-105}{\milli\electronvolt}, & t_h&=\SI{31}{\milli\electronvolt},\\
\mathcal E_{0z}&=\SI{-54}{\milli\electronvolt}, & g_1&=\SI{2}{\milli\electronvolt},  \\
\lambda&=\SI{8.3}{\milli\electronvolt}, & g_2&=\SI{5}{\milli\electronvolt},
\end{array}
\end{equation}
are taken from Ref.~\onlinecite{Vivek2017}. These were obtained by fitting the band structure to experimental photoemission data and were successfully used in Ref.~\onlinecite{Vaz2019}.

The expectation value
\begin{equation}\label{eq:expectation_valueS}
\Braket{\vec s}_{\vec k}^n = \Braket{\Psi_{\vec k}^n |\hat{\vec s} | \Psi_{\vec k} ^n }
\end{equation}
of the spin operator $\hat{\vec s}$ with respect to a Bloch state $\Ket{\Psi_{\vec k} ^n}$ ($n$ band index) is expressed in terms of the Pauli matrices,
\begin{equation}\label{eq:spin_operators}
\hat{s}_i=\hat{\sigma}_i  \otimes \mathds{1}_{4 \times 4}, \quad i = x, y ,z.
\end{equation}

The expectation value 
\begin{equation}\label{eq:expectation_valueL}
\Braket{\vec l}_{\vec k}^n = \Braket{\Psi_{\vec k}^n |\hat{\vec l} | \Psi_{\vec k} ^n }
\end{equation}
of the orbital moment $\hat{\vec l}$ reads 
\begin{equation}
\hat{l}_i = \mathds{1}_{2 \times 2} \otimes  \hat{\lambda}_i \ ,
\end{equation}
with
\begin{align*}
\hat{\lambda}_x & = \begin{pmatrix}
0 & 0 & 0 & -\mathrm i  \\
0 & 0 & 0 & -\mathrm i  \\
0 & 0 & 0 & 0  \\
\mathrm i & \mathrm i & 0 & 0
\end{pmatrix},
\\
\hat{\lambda}_y & = \begin{pmatrix}
0 & 0 & \mathrm i & 0 \\
0 & 0 & \mathrm i & 0  \\
-\mathrm i & -\mathrm i & 0 & 0 \\
0 & 0 & 0 & 0
\end{pmatrix}, 
\\ 
 \hat{\lambda}_z  & = \begin{pmatrix}
0 & 0 & 0 & 0 \\
0 & 0 & 0 & 0  \\
0 & 0 & 0 & \mathrm i  \\
0 & 0 & -\mathrm i & 0
\end{pmatrix}.
\end{align*}

\section{Transport calculations}\label{sec:boltzmann}
The magnetic moment $\vec m_\mathrm{s}$ originating from the spin moment is given by 
\begin{equation}\label{eq:spin_density_s}
\vec m_\mathrm{s} = \frac{g_s \mu_\text B}{\hbar} \sum \limits_{\vec k} f_{\vec k} \Braket{\vec s} _{\vec k}
\end{equation}
in the semiclassical Boltzmann theory for transport utilized here.  Land{\'e}'s g-factor reads $g_s \approx 2$. Here and in the following, the multi-index $\vec k$ comprises the crystal momentum as well as the band index, $\vec k \equiv \left(\hbar \vec k, n\right)$.  Likewise, the orbital moment yields a magnetic moment
\begin{equation}\label{eq:spin_density_l}
\vec m_\mathrm{l} = \frac{g_l \mu_\text B}{\hbar} \sum \limits_{\vec k} f_{\vec k} \Braket{\vec l} _{\vec k}
\end{equation}
with $g_l = 1$.

The distribution function $f_{\vec k} = f_{\vec k}^0 + g_{\vec k}$ is split into an equilibrium part -- that is the Fermi-Dirac distribution function $f_{\vec k}^0$  -- and a nonequilibrium part $g_{\vec k}$. In nonmagnetic systems, $f_{\vec k}^0$ does not contribute to $\vec m_\mathrm{s} $ and $\vec m_\mathrm{l}$.

The Boltzmann equation for a stationary and spatially homogeneous system reads
\begin{equation}\label{eq:Boltzmann}
\dot{\vec k} \frac{\partial f_{\vec k}}{\partial \vec k}=\left(\frac{\partial f_{\vec k}}{\partial t} \right) _\text{scatt}.
\end{equation}
Using the semiclassical equation of motion
\begin{equation}\label{semiclassical_k}
\dot{\vec k}= - \frac{e}{\hbar} \vec E,
\end{equation}
expressing the right hand side in terms of microscopic transition probability rates $P_{\vec k' \gets \vec k}$,
\begin{equation}\label{eq:scattering_term}
\left(\frac{\partial f_{\vec k}}{\partial t} \right)_\text{scatt}= \sum \limits_{\vec k'} \left( P_{\vec k \gets \vec{k}'} g_{\vec k'} - P_{\vec{k}' \gets \vec k} g_{\vec k} \right),
\end{equation}
and with the linear \textit{ansatz}
\begin{equation}\label{eq:linear_ansatz}
g_{\vec k} = \frac{\partial f_{\vec k} ^0}{\partial \mathcal E} e \vec \Lambda_{\vec k} \cdot \vec E,
\end{equation}
the Boltzmann equation is linearized and takes the form
\begin{equation}\label{eq:Boltzmann_linearized}
\vec \Lambda_{\vec k}= \tau_{\vec k} \left( \vec v_{\vec k} + \sum \limits_{\vec k'} P_{\vec k \gets \vec k' } \vec \Lambda_{\vec k'} \right)
\end{equation}
for the mean free path  $\vec \Lambda_{\vec k}$. The momentum relaxation time is given by 
\begin{equation}\label{eq:relaxation_time}
\tau_{\vec k} = \left( \sum \limits_{\vec k'} P_{\vec k' \gets \vec k} \right) ^{-1}.
\end{equation}

For the calculations presented in this Paper we neglect the so-called scattering-in term $\tau_{\vec k} \sum \limits_{\vec k'} P_{\vec k \gets \vec k' } \vec \Lambda_{\vec k'}$ which appears in Eq.~\eqref{eq:Boltzmann_linearized}. With the further assumptions of zero temperature and a constant relaxation time $\tau_{\vec k} = \tau_0$, Eqs.~\eqref{eq:induced_m} and \eqref{eq:chi_boltzmann} yield the magnetic moment induced by an external electric field within the linear-response regime.

\end{document}